\documentclass[12pt,a4paper]{article}

\usepackage[utf8]{inputenc}
\usepackage{amsmath}
\usepackage{url}
\usepackage{mathtools}

\begin{document}

\title{Constructing Error-Correcting Binary Codes \\ Using Transitive Permutation Groups}

\author{Antti~Laaksonen, Patric Östergård}

\maketitle

\begin{abstract}
Let $A_2(n,d)$ be the maximum size of a binary code
of length $n$ and minimum distance $d$.
In this paper we present the following new lower bounds:
$A_2(18,4) \ge 5632$, $A_2(21,4) \ge 40960$,
$A_2(22,4) \ge 81920$, $A_2(23,4) \ge 163840$, $A_2(24,4) \ge 327680$,
$A_2(24,10) \ge 136$, and $A_2(25,6) \ge 17920$.
The new lower bounds are a result of a systematic computer
search over transitive permutation groups.
\end{abstract}

\section{Introduction}

A binary code $C$ of length $n$ is
a set of binary vectors of length $n$.
Each vector $w=(w_1,w_2,\ldots,w_n)$
in the code is called a codeword.
The minimum distance of $C$ is the minimum of $d_H(a,b)$
where $a,b \in C$ and $d_H$ is the Hamming distance, i.e.,
the number of coordinates where $a$ and $b$ differ.
The size of $C$ is the number of codewords that it contains.

Let $A_2(n,d)$ be the maximum size of a binary code
of length $n$ and minimum distance $d$.
A long-standing problem in information theory has been to find
lower and upper bounds for $A_2(n,d)$ values \cite{BBMOS78}.
In this paper we present several improved lower bounds
that are summarized in Table \ref{tab1}.

\begin{table}[h]
\renewcommand{\arraystretch}{1.3}
\caption{New Lower Bounds for Code Sizes}
\label{tab1}
\centering
\begin{tabular}{|cc|}
\hline
Old Lower Bound & New Lower Bound \\
$A_2(18,4) \ge 5312$ \cite{CHLL97} & $A_2(18,4) \ge 5632$ \\
$A_2(21,4) \ge 36864$ \cite{BBMOS78} & $A_2(21,4) \ge 40960$ \\
$A_2(22,4) \ge 73728$ \cite{BBMOS78} & $A_2(22,4) \ge 81920$ \\
$A_2(23,4) \ge 147456$ \cite{BBMOS78} & $A_2(23,4) \ge 163840$ \\
$A_2(24,4) \ge 294912$ \cite{BBMOS78} & $A_2(24,4) \ge 327680$ \\
$A_2(24,10) \ge 128$ \cite{BBMOS78} & $A_2(24,10) \ge 136$ \\
$A_2(25,6) \ge 16384$ \cite{BBMOS78} & $A_2(25,6) \ge 17920$ \\
\hline
\end{tabular}
\end{table}

Our new lower bounds are a result of a systematic computer search
over transitive permutation groups.
We focus on codes that consist of orbits of codewords
generated by a permutation group,
and reduce the problem of finding a maximum size code
into a maximum weight clique problem.

The motivation for the approach is that several existing codes have this kind of
symmetric structure, which suggests that it is a good way to design codes.
In addition,
complete databases of transitive permutation groups
are available, which makes an automatic computer search possible.

\section{Code Construction}

We represent a codeword $w$ of length $n$ as a doubleword $x$
that is a binary vector $(x_1,x_2\ldots,x_{2n})$ of length $2n$.
The first $n$ elements of $x$
correspond to the actual codeword,
and the last $n$ elements of $x$
are the inverse of the codeword.
More precisely, $x_i=w_i$ for $i=1,2,\ldots,n$,
and $x_i=1-w_{i-n}$ for $i=n+1,n+2,\ldots,2n$.

Let $G$ be a permutation group of degree $2n$.
For any doubleword $x$ of length $2n$,
we use the notation $G.x$ for the orbit of $x$, i.e.,
the set $G.x$ contains all doublewords that can be
generated from $x$ using the permutations in $G$.
We construct codes that consist of
one or more orbits, i.e., each code is of form $G.x_1 \cup G.x_2 \cup \cdots \cup G.x_k$
where $x_1,x_2,\ldots,x_k$ are the corresponding
orbit representatives.

The benefit of using the doubleword representation
is that permutations of both coordinates
and coordinate values in the original codeword can
be expressed as permutations of coordinates in the doubleword.
To allow only this type of permutations,
we require that for each permutation $p \in G$,
$|p(i)-p(n+i)|=n$ for $i=1,2,\ldots,n$.

The construction of a code of minimum distance $d$
can be seen as a graph problem.
Let $d_H(x)$ be the minimum distance of
two codewords in $G.x$,
and let $d_H(x,y)$ be the minimum distance
of any codewords $a \in G.x$ and $b \in G.y$.
Let us construct a graph whose nodes are
the orbit representatives $x$ such that $d_H(x) \ge d$,
and there is an edge between nodes $x$ and $y$
if $d_H(x,y) \ge d$.
Now any clique in the graph corresponds
to a code of minimum distance $d$,
and the weight of the clique
is equal to the size of the code.

Using this approach, new codes can be discovered
using a computer search.
We constructed a large number of potential permutation
groups based on databases of transitive permutation groups.
All transitive permutation groups up to degree 30
are available in GAP \cite{GAP}, and extensions
up to degree 47 can be obtained from \cite{CH08,Hul05}.

After fixing the permutation group,
the remaining problem is to select the orbits
that will be included in the code.
The problem of constructing a maximum size code
is equal to the problem of finding a maximum weight
clique in the corresponding graph.
Even if the problem itself is NP-hard,
heuristic methods can yield good results.
We used the Cliquer \cite{NO03} tool for searching
for maximum weight cliques.

\section{New Results}

In this section we present the codes
that correspond to the new lower bounds.
For each code we give the generators
of the permutation group,
and the orbit representatives from which
the code can be generated.

\subsection{Lower Bound $A_2(18,4) \ge 5632$}

Let

\begin{center}
\begin{tabular}{rcl}
$G$ & $=\{$ & 
$\begin{psmallmatrix}2 & 5 & 3 & 4 & 19 & 22 & 20 & 21\end{psmallmatrix}
\begin{psmallmatrix}6 & 34 & 7 & 33 & 23 & 17 & 24 & 16\end{psmallmatrix}$ \\
& & $\begin{psmallmatrix}8 & 15 & 9 & 14 & 25 & 32 & 26 & 31\end{psmallmatrix}
\begin{psmallmatrix}10 & 13 & 28 & 29 & 27 & 30 & 11 & 12\end{psmallmatrix}$, \\
& & $\begin{psmallmatrix}2 & 34 & 3 & 33 & 19 & 17 & 20 & 16\end{psmallmatrix}
\begin{psmallmatrix}4 & 14 & 22 & 32 & 21 & 31 & 5 & 15\end{psmallmatrix}$ \\
& & $\begin{psmallmatrix}6 & 12 & 7 & 13 & 23 & 29 & 24 & 30\end{psmallmatrix}
\begin{psmallmatrix}8 & 27 & 9 & 11 & 25 & 10 & 26 & 28\end{psmallmatrix}$, \\
& & $\begin{psmallmatrix}2 & 33 & 28 & 9 & 19 & 16 & 11 & 26\end{psmallmatrix}
\begin{psmallmatrix}3 & 34 & 27 & 8 & 20 & 17 & 10 & 25\end{psmallmatrix}$ \\
& & $\begin{psmallmatrix}4 & 15 & 13 & 7 & 21 & 32 & 30 & 24\end{psmallmatrix}
\begin{psmallmatrix}5 & 14 & 12 & 23 & 22 & 31 & 29 & 6\end{psmallmatrix}$, \\
& & $\begin{psmallmatrix}2 & 6 & 3 & 24 & 19 & 23 & 20 & 7\end{psmallmatrix}
\begin{psmallmatrix}4 & 34 & 22 & 16 & 21 & 17 & 5 & 33\end{psmallmatrix}$ \\
& & $\begin{psmallmatrix}8 & 29 & 26 & 30 & 25 & 12 & 9 & 13\end{psmallmatrix}
\begin{psmallmatrix}10 & 32 & 28 & 14 & 27 & 15 & 11 & 31\end{psmallmatrix} \}$ \\
\end{tabular}
\end{center}

and

\begin{center}
\begin{tabular}{rcl}
$X$ & $=\{$ &
\scriptsize{1000111111111111101110000000000000}, \\
& & \scriptsize{0010011111111111111011000000000000}, \\
& & \scriptsize{0100101111111111110110100000000000}, \\
& & \scriptsize{1011001111111111101001100000000000}, \\
& & \scriptsize{0111010111111111110001010000000000}, \\
& & \scriptsize{0101011011111111110101001000000000}, \\
& & \scriptsize{1001110011111111101100011000000000}, \\
& & \scriptsize{1110110011111111100010011000000000}, \\
& & \scriptsize{0000010011111111111111011000000000}, \\
& & \scriptsize{0101110101111111110100010100000000}, \\
& & \scriptsize{1000000001111111101111111100000000}, \\
& & \scriptsize{1001011010111111101101001010000000}, \\
& & \scriptsize{0011011010011111111001001011000000} \normalsize{$\}.$}
\end{tabular}
\end{center}

The group $G$ has degree 34, coordinates 1 and 18
are fixed and other coordinates are permuted
transitively.
The code generated by $G$ from $X$ has length 17,
size 5632 and minimum distance 3.

\subsection{Lower Bound $A_2(24,4) \ge 327680$}

Let

\begin{center}
\begin{tabular}{rcl}
$G$ & $=\{$ &
$\begin{psmallmatrix}1 & 7 & 33 & 3 & 5 & 35 & 26 & 32 & 10 & 28 & 30 & 12\end{psmallmatrix}$ \\
& & $\begin{psmallmatrix}2 & 8 & 34 & 4 & 6 & 36 & 25 & 31 & 9 & 27 & 29 & 11\end{psmallmatrix}$ \\
& & $\begin{psmallmatrix}13 & 19 & 45 & 15 & 17 & 47 & 38 & 44 & 22 & 40 & 42 & 24\end{psmallmatrix}$ \\
& & $\begin{psmallmatrix}14 & 20 & 46 & 16 & 18 & 48 & 37 & 43 & 21 & 39 & 41 & 23\end{psmallmatrix},$ \\
& & $\begin{psmallmatrix}1 & 35\end{psmallmatrix}
\begin{psmallmatrix}2 & 36\end{psmallmatrix}
\begin{psmallmatrix}3 & 9 & 27 & 33\end{psmallmatrix}
\begin{psmallmatrix}4 & 34 & 28 & 10\end{psmallmatrix}$ \\
& & $\begin{psmallmatrix}5 & 8\end{psmallmatrix}
\begin{psmallmatrix}6 & 7\end{psmallmatrix}
\begin{psmallmatrix}11 & 25\end{psmallmatrix}
\begin{psmallmatrix}12 & 26\end{psmallmatrix}
\begin{psmallmatrix}13 & 47\end{psmallmatrix}
\begin{psmallmatrix}14 & 48\end{psmallmatrix} $ \\
& & $\begin{psmallmatrix}15 & 21 & 39 & 45\end{psmallmatrix}
\begin{psmallmatrix}16 & 46 & 40 & 22\end{psmallmatrix}
\begin{psmallmatrix}17 & 20\end{psmallmatrix}
\begin{psmallmatrix}18 & 19\end{psmallmatrix} $ \\
& & $
\begin{psmallmatrix}24 & 38\end{psmallmatrix}
\begin{psmallmatrix}29 & 32\end{psmallmatrix}
\begin{psmallmatrix}30 & 31\end{psmallmatrix}
\begin{psmallmatrix}41 & 44\end{psmallmatrix}
\begin{psmallmatrix}42 & 43\end{psmallmatrix} \}$ \\
\end{tabular}
\end{center}

and

\begin{center}
\begin{tabular}{rcl}
$X$ & $=\{$ &
\scriptsize{000101111111111111111111111010000000000000000000}, \\
& & \scriptsize{000000010111111111111111111111101000000000000000}, \\
& & \scriptsize{001100100111111111111111110011011000000000000000}, \\
& & \scriptsize{110000100111111111111111001111011000000000000000}, \\
& & \scriptsize{010110010101111111111111101001101010000000000000}, \\
& & \scriptsize{101010010101111111111111010101101010000000000000}, \\
& & \scriptsize{111111011111011111111111000000100000100000000000}, \\
& & \scriptsize{001001011111011111111111110110100000100000000000}, \\
& & \scriptsize{100101001111011111111111011010110000100000000000}, \\
& & \scriptsize{011000110111011111111111100111001000100000000000}, \\
& & \scriptsize{010100000111011111111111101011111000100000000000}, \\
& & \scriptsize{110101010011011111111111001010101100100000000000}, \\
& & \scriptsize{111010010011011111111111000101101100100000000000}, \\
& & \scriptsize{101001000011011111111111010110111100100000000000}, \\
& & \scriptsize{100001010101011111111111011110101010100000000000}, \\
& & \scriptsize{000000000001011111111111111111111110100000000000} \normalsize{$\}.$}
\end{tabular}
\end{center}

The group $G$ consists of two copies of
a transitive permutation group of degree $24$.
The code generated by $G$ from $X$ has length 24,
size 327680 and minimum distance 4.
Since $A_2(n-1,d) \ge A_2(n,d)/2$,
we also get lower bounds $A_2(23,4) \ge 163840$,
$A_2(22,4) \ge 81920$, and $A_2(21,4) \ge 40960$.

\subsection{Lower Bound $A_2(24,10) \ge 136$}

Let

\begin{center}
\begin{tabular}{rcl}
$G$ & $=\{$ &
$\begin{psmallmatrix}1 & 27 & 30 & 12\end{psmallmatrix}
\begin{psmallmatrix}2 & 28 & 31 & 10\end{psmallmatrix}
\begin{psmallmatrix}3 & 6 & 36 & 25\end{psmallmatrix}$ \\
& & $\begin{psmallmatrix}4 & 7 & 34 & 26\end{psmallmatrix}
\begin{psmallmatrix}5 & 8 & 35 & 33\end{psmallmatrix}
\begin{psmallmatrix}9 & 29 & 32 & 11\end{psmallmatrix}$ \\
& & $\begin{psmallmatrix}13 & 39 & 42 & 24\end{psmallmatrix}
\begin{psmallmatrix}14 & 40 & 43 & 22\end{psmallmatrix}
\begin{psmallmatrix}15 & 18 & 48 & 37\end{psmallmatrix}$ \\
& & $\begin{psmallmatrix}16 & 19 & 46 & 38\end{psmallmatrix}
\begin{psmallmatrix}17 & 20 & 47 & 45\end{psmallmatrix}
\begin{psmallmatrix}21 & 41 & 44 & 23\end{psmallmatrix}$, \\
& & $\begin{psmallmatrix}1 & 36 & 26 & 33 & 11 & 31\end{psmallmatrix}
\begin{psmallmatrix}2 & 9 & 35 & 7 & 25 & 12\end{psmallmatrix}$ \\
& & $\begin{psmallmatrix}3 & 4 & 6 & 29 & 10 & 32\end{psmallmatrix}
\begin{psmallmatrix}5 & 34 & 8 & 27 & 28 & 30\end{psmallmatrix}$ \\
& & $\begin{psmallmatrix}13 & 48 & 38 & 45 & 23 & 43\end{psmallmatrix}
\begin{psmallmatrix}14 & 21 & 47 & 19 & 37 & 24\end{psmallmatrix}$ \\
& & $\begin{psmallmatrix}15 & 16 & 18 & 41 & 22 & 44\end{psmallmatrix}
\begin{psmallmatrix}17 & 46 & 20 & 39 & 40 & 42\end{psmallmatrix} \}$
\end{tabular}
\end{center}

and

\begin{center}
\begin{tabular}{rcl}
$X$ & $=\{$ &
\scriptsize{100101010010000111111111011010101101111000000000}, \\
& & \scriptsize{001000101000100010011111110111010111011101100000}, \\
& & \scriptsize{010011100100110011101111101100011011001100010000}, \\
& & \scriptsize{000110101111000110101111111001010000111001010000}, \\
& & \scriptsize{001010110010101100100111110101001101010011011000}, \\
& & \scriptsize{010010111001011101101101101101000110100010010010} \normalsize{$\}.$}
\end{tabular}
\end{center}

The group $G$ consists of two copies of
a transitive permutation group of degree $24$.
The code generated by $G$ from $X$ has length 24,
size 136 and minimum distance 10.

\subsection{Lower Bound  $A_2(25,6) \ge 17920$}

Let

\begin{center}
\begin{tabular}{rcl}
$G$ & $=\{$ &
$\begin{psmallmatrix}1 & 26\end{psmallmatrix}
\begin{psmallmatrix}2 & 25\end{psmallmatrix}
\begin{psmallmatrix}3 & 4 & 27 & 28\end{psmallmatrix}
\begin{psmallmatrix}5 & 30 & 29 & 6\end{psmallmatrix}$ \\
& & $\begin{psmallmatrix}7 & 32\end{psmallmatrix}
\begin{psmallmatrix}8 & 31\end{psmallmatrix}
\begin{psmallmatrix}9 & 12 & 33 & 36\end{psmallmatrix}
\begin{psmallmatrix}10 & 35\end{psmallmatrix}
\begin{psmallmatrix}11 & 34\end{psmallmatrix}$ \\
& & $\begin{psmallmatrix}13 & 38\end{psmallmatrix}
\begin{psmallmatrix}14 & 37\end{psmallmatrix}
\begin{psmallmatrix}15 & 16 & 39 & 40\end{psmallmatrix}
\begin{psmallmatrix}17 & 42 & 41 & 18\end{psmallmatrix}$ \\
& & $\begin{psmallmatrix}19 & 44\end{psmallmatrix}
\begin{psmallmatrix}20 & 43\end{psmallmatrix}
\begin{psmallmatrix}21 & 24 & 45 & 48\end{psmallmatrix}
\begin{psmallmatrix}22 & 47\end{psmallmatrix}
\begin{psmallmatrix}23 & 46\end{psmallmatrix}$, \\
& & $\begin{psmallmatrix}1 & 31 & 12 & 25 & 7 & 36\end{psmallmatrix}
\begin{psmallmatrix}2 & 29 & 35\end{psmallmatrix}
\begin{psmallmatrix}3 & 32 & 10 & 27 & 8 & 34\end{psmallmatrix}$ \\
& & $\begin{psmallmatrix}4 & 6 & 33 & 28 & 30 & 9\end{psmallmatrix}
\begin{psmallmatrix}5 & 11 & 26\end{psmallmatrix}
\begin{psmallmatrix}13 & 43 & 24 & 37 & 19 & 48\end{psmallmatrix}$ \\
& & $\begin{psmallmatrix}14 & 41 & 47\end{psmallmatrix}
\begin{psmallmatrix}15 & 44 & 22 & 39 & 20 & 46\end{psmallmatrix}$ \\
& & $\begin{psmallmatrix}16 & 18 & 45 & 40 & 42 & 21\end{psmallmatrix}
\begin{psmallmatrix}17 & 23 & 38\end{psmallmatrix} \}$
\end{tabular}
\end{center}

and

\begin{center}
\begin{tabular}{rcl}
$X$ & $=\{$ &
\scriptsize{000110000100111111111111111001111011000000000000}, \\
& & \scriptsize{000110111101001111111111111001000010110000000000}, \\
& & \scriptsize{010011001001011111111111101100110110100000000000}, \\
& & \scriptsize{001101001110011111111111110010110001100000000000}, \\
& & \scriptsize{001001010110110111111111110110101001001000000000}, \\
& & \scriptsize{000001111000011111111111111110000111100000000000}, \\
& & \scriptsize{000010110010110111111111111101001101001000000000}, \\
& & \scriptsize{101110111110111111111111010001000001000000000000}, \\
& & \scriptsize{101101110111001111111111010010001000110000000000}, \\
& & \scriptsize{111001000010001111111111000110111101110000000000}, \\
& & \scriptsize{111110000111011111111111000001111000100000000000}, \\
& & \scriptsize{110010110001011111111111001101001110100000000000}, \\
& & \scriptsize{110010001011110111111111001101110100001000000000}, \\
& & \scriptsize{111101001101110111111111000010110010001000000000}, \\
& & \scriptsize{110110100110110111111111001001011001001000000000} \normalsize{$\}.$}
\end{tabular}
\end{center}

The group $G$ consists of two copies of
a transitive permutation group of degree $24$.
The code generated by $G$ from $X$ has length 24,
size 17920 and minimum distance 5.

\end{document}